\documentclass[floats, aps]{revtex4}
\usepackage{amsmath}
\usepackage{amssymb}
\usepackage{mathrsfs}
\usepackage{graphicx}
\usepackage{enumerate}
\begin{document}
  \title{Multiple Core-Hole Coherence in X-Ray Four-Wave-Mixing Spectroscopies}
  \author{Shaul Mukamel}
  \affiliation{Department of Chemistry, \\
    University of California, Irvine \\
    Irvine, CA  92697-2025}
  \email{:smukamel@uci.edu}
  \altaffiliation{\\ Submitted to Phys. Rev. B}
  \begin{abstract}
    Correlation-function expressions are derived for the coherent
    nonlinear response of molecules to three resonant ultrafast pulses
    in the x-ray regime. The ability to create two-core-hole states
    with controlled attosecond timing in four-wave-mixing and pump
    probe techniques should open up new windows into the response of
    valence electrons, which are not available from incoherent x-ray
    Raman and fluorescence techniques.  Closed expressions for the
    necessary four-point correlation functions are derived for the
    electron-boson model by using the second order cumulant expansion
    to describe the fluctuating potentials. The information obtained
    from multidimensional nonlinear techniques could be used to test
    and refine this model, and establish an anharmonic oscillator
    picture for electronic excitations.

  \end{abstract}
  \maketitle

  %-------------------------------------------------------
  %-------------------------------------------------------
  %-------------------------------------------------------
  \section{Introduction}
  %-------------------------------------------------------
  %-------------------------------------------------------
  %-------------------------------------------------------

  The development of bright attosecond soft and hard x-ray sources
  such as the free electron laser has triggered considerable interest
  in all-x-ray nonlinear spectroscopy\cite{murnane91,krausz,
  cherguireview, cherguimukamelpreface}.  In resonant optical
  techniques in the visible, a photon is tuned to high frequency
  ($\sim 2$eV) electronic transitions, but a wealth of information is
  provided on lower-frequency ($< 0.4$eV) nuclear (vibrational and
  phonon) degrees of freedom which are accessible through multiphoton
  (e.g. Raman-type) resonances with differences (and higher
  combinations) of visible photons.  In a completely analogous manner,
  combinations of x-ray photons resonant with high frequency (~keV)
  core transitions can probe the lower frequency ($< 50$eV) valence
  electronic excitations. By exploiting this analogy, we can use the
  theoretical apparatus developed for probing electronic and
  vibrational coherences in nonlinear optics\cite{mukbook}, to predict
  nonlinear x-ray signals and design new multiple pulse experiments.
  For example, the concepts underlying multidimensional
  techniques\cite{smannreview} which provide extremely valuable
  information on optical excitations in molecular aggregates can be
  extended to probe correlations among multiple core hole states.

  Nozieres and De Dominicis had proposed a model Hamiltonian for
  resonant x-ray processes in metals\cite{nozieres69}. The $\sim
  \omega^{\alpha}$ threshold (Fermi edge singularity) behavior of
  x-ray absorption was calculated.  The interplay of the large number
  of electron hole pairs and the vanishing of the many-electron
  wavefunction overlap with and without the core hole (Anderson's
  Catastrophe) can result in either a diverging ($\alpha<0$) or
  converging ($\alpha > 0$) lineshape at threshold, as shown by
  Mahan\cite{mahan82, hedin96, nordgren97}.  Signatures of this
  singularity in femtosecond optical pump probe spectroscopy of doped
  semiconductor nanostructures have been studied.\cite{perakis} Higher
  order radiative processes such as spontaneous emission (fluorescence
  or Raman) or nonlinear wave mixing provide more detailed insights
  through multipoint correlation functions\cite{nozieresandabrams,
  beyondoneelectronmodel, mukbook, agrenreview99, agren01} .  The
  formal theory of x-ray response closely resembles its visible or
  infrared counterpart where the relevant correlation and response
  functions and possible time orderings have been studied
  extensively.\cite{mukxraypaper} It is also closely connected with
  the treatment of currents in open molecular systems coupled to
  electrodes.\cite{upendra}

  In this paper we apply the density matrix Liouville space
  formalism\cite{mukbook} to calculate time resolved x-ray four-wave
  mixing signals and compare them with spontaneous emission and pump
  probe spectroscopy.  Much current activity is focused on the
  application of time resolved diffraction to probe structural changes
  such as surface melting.\cite{melting} From a theoretical point
  these can be described using the existing formalism of diffraction
  by simply including the parametric time dependence of the electronic
  charge density; coherence does not play a role in these
  techniques. Other experiments have been carried out using a visible
  or an infrared pump followed by the absorption of an x-ray
  probe. These techniques provide x-ray snapshots of vibrational
  coherence or coherence among valence electronic
  states. \cite{agren05} Photoelectron spectroscopy provides
  additional novel ultrafast probes.\cite{agren04, gelmukhanov05} 

  Pure resonant x-ray nonlinear optics of the type considered here can
  probe high frequency coherence of many-electron states involving
  core hole transitions and provide multidimensional real space
  pictures for the response of valence electrons to external
  perturbations.  The necessary multipoint correlation functions may
  be calculated using several levels of theory. (i) Multiple
  summations over the many electron states of the valence system with
  $N$, $N+1$, and $N+2$ electrons in the presence of zero, one, and
  two core holes respectively. (ii) The transition potential method,
  which uses a reference system with partially filled orbitals
  \cite{degroottransstate, nexafsofadsorbates,dftinnershell,
  degroot74} (iii) many-body Green function perturbative
  techniques. \cite{nozieresandabrams, agren01,
  beyondoneelectronmodel, hedin99, hedin96, nordgren97, cederbaum81}
  (iv) Replace the original Hamiltonian by the electron-boson model
  (EBM) and treat charge-density fluctuations as classical
  oscillations. \cite{langreth, hedin99, langreth70,lundqvist67,
  overhauser71}

  Method (i) is conceptually the simplest but the most expensive
  numerically.  The sum over state (SOS) expressions for nonlinear
  response functions allow us to employ any level of quantum chemistry
  for computing electronically excited states. TDDFT, for example,
  provides a relatively efficient way for computing a large number of
  electronically excited states.\cite{gross} The relevant states are
  determined by the pulse bandwidth. e.g. $3.75$eV for a $10$as
  pulse. Method (ii) Represents systems with different numbers of core
  holes by varying the occupation numbers of a single set of reference
  orbitals. This approximate method works well for core level
  spectroscopies of small molecules, and may be extended to the
  nonlinear response. Method (iii) was used by Nozieres and coworkers
  to compute XANES and x-ray Raman
  spectra.\cite{nozieres69,nozieresandabrams} It is formally exact,
  and allows the development of powerful approximations. The EBM is by
  far the simplest to implement, since it is exactly solvable by the
  second order cumulant expansion. The model is not expected to apply
  for small molecules where the core electrons added to the valence
  band by the absorption of an x-ray photon need to be treated
  explicitly.  It has been tested and found to work quite well for
  solids (semiconductors and metals). Plasmon satellites have been
  predicted in x-ray photoemission.\cite{hedin85, hedin96,
  beyondoneelectronmodel} and inelastic scattering.\cite{schulke} They
  are less clearly seen in x-ray absorption, but have indirect
  signatures in e.g. the oscillator strengths. Using the EBM,
  electronic excitations are described as anharmonic oscillations
  whose parameters may be extracted from coherent nonlinear x-ray
  techniques.

  %-------------------------------------------------------
  %-------------------------------------------------------
  %-------------------------------------------------------
  \section{The Nonlinear X-Ray Response}
  %-------------------------------------------------------
  %-------------------------------------------------------
  %------------------------------------------------------- 
  We start with the Mahan-Nozieres-De Dominicis(MND)
  Hamiltonian\cite{mahan82,nozieres69, almbladh01, hedin99, hedin85}
  \begin{equation}
    H = \sum_k \epsilon_k c_k^{\dagger} c_k + \sum_n \epsilon_n
    d_n^{\dagger}d_n + \sum_{\substack{k k' \\ n}} V_{k k', n}
    c_k^{\dagger} c_{k'} d_n d_n^{\dagger}
  \end{equation}
  where $c_k^{\dagger}$, $d_n^{\dagger}$ are the Fermi creation
  operators for a valence and a core electron respectively,
  $\epsilon_n$ is the energy of core state $n$, and $V_{k k', n}$ is
  the potential acting on the valence electrons due to the $n$th core
  hole.

  The dipole interaction with the x-ray field in the rotating wave
  approximation (RWA) is
  \begin{equation}
    H_{int} = \sum_n \left[ E(\mathbf{r},t) B_n^{\dagger} + E^*(\mathbf{r},t) B_n\right]
  \end{equation}
  where 
  \begin{equation}
    B^{\dagger}_n = \sum_k \mu_{kn} c_k^{\dagger} d_n,\qquad B_n =
    \sum_k \mu_{nk} c_k d^{\dagger}_n
  \end{equation}
  are creation (annihilation) operators for core-hole
  excitons. $\mu_{kn}$ is the dipole matrix element between the $n$'th core
  orbital and the $k$'th valence orbital, and $E(\mathbf{r},t)$ is the 
  complex field envelope (see Eq. (\ref{eq:optlaukas})). The possible
  transitions between the zero, one and two core hole states are shown
  in Fig. 1.
  
  The $j$'th order molecular response to the x-ray field is described by
  the induced polarization~\cite{mukbook}:
  %---------------------------------------------------------
  \begin{equation}\label{eq:polarizationtrcdef}
    P^{(j)}(\mathbf{r},t) \equiv Tr( \hat{P}
    \hat{\rho}^{(j)}(t) )
  \end{equation}
  %--------------------------------------------------------- 
  where $Tr(\ldots)$ denotes the trace and
  \begin{equation}\label{eq:polarizationdef}
    \hat{P} = \sum_n (B_n + B^{\dagger}_n)
  \end{equation}
  is the polarization operator. $\hat{\rho}^{(j)}(t)$ is the density
  matrix describing the state of the molecule obtained by solving the
  Liouville equation to $j$'th order in the field.
  %---------------------------------------------------------
  \begin{equation}
    \frac{d\hat{\rho}}{d t}=-i[\hat{H},\hat{\rho}],
  \end{equation}
  %---------------------------------------------------------
  where the total Hamiltonian $\hat{H}=\hat{H}_0 +\hat{H}_{int}$,
  $\hat{H}_0$ is the material Hamiltonian.  

  Eq. (\ref{eq:polarizationtrcdef}) can be expanded as
  %---------------------------------------------------------
  \begin{eqnarray}
    P^{(j)}(\mathbf{r}, t ) &=
    &\int_0^{\infty}d t_n
    \int_0^{\infty}d t_{n-1}
    \ldots
    \int_0^{\infty}d t_1
    S^{(j)}( t_n, t_{n-1},\ldots, t_1)
    \times \nonumber\\
    &&              E(\mathbf{r},t-t_n)
    E(\mathbf{r},t-t_n-t_{n-1}) \ldots
    E(\mathbf{r},t-t_n-t_{n-1}\ldots-t_1).
  \end{eqnarray}
  %---------------------------------------------------------
  where $S^{(j)}( t_n, t_{n-1},\ldots,t_1)$ denotes the
  $j$--th order \emph{response function} and $t_n
  \equiv \tau_{n+1}-\tau_n$ is the time delay between two
  consecutive interactions with the x-ray field (see Fig. 2).

  %--------------------------------------------------------- 
  The first order polarization is related to the linear response
  function $S^{(1)}$:
  %---------------------------------------------------------
  \begin{equation}
    P^{(1)}(\mathbf{r},t) = \int_0^{\infty}dt_1 S^{(1)}(t_1)
    E(\mathbf{r},t-t_1),
  \end{equation}
  %--------------------------------------------------------- 
  $S^{(1)}$ is a second rank tensor with respect to the polarization
  direction. For clarity we do not use tensor notation.

  Here
  %--------------------------------------------------------------
  \begin{equation}\label{eq:sdefusingj}
    S^{(1)}(t_1)=i\theta(t_1)[J(t_1)-J^{\ast}(t_1)],
  \end{equation}
  %--------------------------------------------------------------
  $J(t_1)=\langle \hat{P}(t_1) \hat{P}(0) \rangle$ is a two point
  correlation function and the polarization operator is given in the
  Heisenberg representation:
  %---------------------------------------------------------
  \begin{equation}
    \hat{P}(t_1)=\exp(i\hat{H}_0 t_1)\hat{P}\exp(-i\hat{H}_0 t_1),
  \end{equation}
  %---------------------------------------------------------
  % Then
  %---------------------------------------------------------
  %\begin{equation}\label{eq:2pointcorr}
  %J(t_1)=.
  %\end{equation}
  %---------------------------------------------------------
  $\theta(t)$ is the Heavyside function ($\theta(t)=0$ for $t<0$,
  $\theta(t)=1$ for $t \geq 0$) which represents causality, and the
  angular brackets $\langle\cdots\rangle$ denote the trace over the
  ground state density matrix:

  The third order polarization is given by
  %---------------------------------------------------------
  \begin{eqnarray}
    P^{(3)}(\mathbf{r},t) &=& 
    \int \!\!\! \int \!\!\! \int_0^{\infty} dt_3 dt_2 dt_1
    %         \int_0^{\infty}dt_3
    %         \int_0^{\infty}dt_2
    %         \int_0^{\infty}dt_1
    S^{(3)}(t_3,t_2,t_1) \times \nonumber \\
    &&{E}(\mathbf{r},t-t_3)
    {E}(\mathbf{r},t-t_3-t_2)
    {E}(\mathbf{r},t-t_3-t_2-t_1).
  \end{eqnarray}
  %---------------------------------------------------------
  The third order response function is similarly given by a sum of
  eight terms, each representing a distinct Liouville space
  pathway~\cite{mukbook}:
  %---------------------------------------------------------
  \begin{equation}\label{eq:response-function-diagonal}
    S(t_3,t_2,t_1) = i^3 \theta(t_3)\theta(t_2)\theta(t_1)\sum_{p
      =1}^{4} \Big[ R_{p}(t_3,t_2,t_1) - R_{p}^{\ast}(t_3,t_2,t_1) \Big],
  \end{equation}
  %---------------------------------------------------------
  where
  %---------------------------------------------------------
  \begin{eqnarray}\label{eq:RandF}
    R_1(t_3,t_2,t_1) & = & F(t_1,t_1+t_2,t_1+t_2+t_3,0), \nonumber \\
    R_2(t_3,t_2,t_1) & = & F(0,t_1+t_2,t_1+t_2+t_3,t_1), \nonumber \\
    R_3(t_3,t_2,t_1) & = & F(0,t_1,t_1+t_2+t_3,t_1+t_2), \nonumber \\
    R_4(t_3,t_2,t_1) & = & F(t_1+t_2+t_3,t_1+t_2,t_1,0),
  \end{eqnarray}
  %---------------------------------------------------------
  and the four-point correlation function is given by:
  %---------------------------------------------------------
  \begin{equation}\label{eq:fourpointCfunction}
    F ( \tau_4,\tau_3,\tau_2,\tau_1) = \langle \hat{P}(\tau_4) \hat{P}(\tau_3)
    \hat{P}(\tau_2) \hat{P}(\tau_1)  \rangle.
  \end{equation}
  %---------------------------------------------------------

  The polarization may be alternatively expressed in the frequency
  domain using the susceptibility tensors:
  $\chi^{(1)}(-\omega_a;\omega_a) \equiv
  \chi^{(1)}(\omega_a) $ and
  $\chi^{(3)}(-\omega_s;\omega_a,\omega_b,\omega_c)$:
  %-----------------------------------------------------------------
  \begin{eqnarray}\label{eq:poliarizacija1}
    P^{(1)}(\mathbf{r},t) = \int_{-\infty}^{\infty}
    d\omega_a \exp(-i\omega_a t) \chi^{(1)}(\omega_a)
    \mathcal{E}(\mathbf{r},\omega_a).
  \end{eqnarray}
  %-----------------------------------------------------------------
  \begin{eqnarray}\label{eq:poliarizacija}
    P^{(3)}(\mathbf{r},t) &=& \int
    \!\!\! \int \!\!\! \int \!\!\! \int_{-\infty}^{\infty} d\omega_s d\omega_a d\omega_b
    d\omega_c
    \exp(-i\omega_s t) \times \nonumber \\
    &&\times
    \chi^{(3)}(-\omega_s;\omega_a,\omega_b,\omega_c)
    \mathcal{E}(\mathbf{r},\omega_a)
    \mathcal{E}(\mathbf{r},\omega_b)
    \mathcal{E}(\mathbf{r},\omega_c).
  \end{eqnarray}
  %-----------------------------------------------------------------
  $\mathcal{E}(\mathbf{r},\omega_a)$ is the x-ray field
  in the frequency domain:
  %------------------------------------------------------------------
  \begin{equation}\label{eq:laukasFurje}
    {\mathcal{E}}(\mathbf{r},\omega)\equiv \int_{-\infty}^{\infty} d\tau
    {E}(\mathbf{r},\tau) \exp(i\omega \tau).
  \end{equation}
  %------------------------------------------------------------------

  The response functions and the nonlinear susceptibilities are
  related by a Fourier transform:
  %----------------------------------------------------------------
  \begin{eqnarray}
    \chi^{(1)}(\omega_a) \equiv
    \int_{0}^{\infty}  dt_1
    S^{(1)}(t_1) \exp(i\omega_a t_1 ),
  \end{eqnarray}
  %----------------------------------------------------------------
  \begin{eqnarray}
    \chi^{(3)}(-\omega_s;\omega_a,\omega_b,\omega_c)
    \equiv
    \frac{1}{3!}\sum_p
    \int  \!\!\!
    \int  \!\!\!
    \int_{0}^{\infty}  dt_3 dt_2 dt_1
    S^{(3)}(t_3,t_2,t_1)\\\nonumber
    \exp(i(\omega_a +\omega_b +\omega_c) t_3 + i(\omega_a
    +\omega_b)t_2 + i\omega_a t_1 ),
  \end{eqnarray}
  %----------------------------------------------------------------
  where $\omega_s\equiv\omega_a +\omega_b +\omega_c$ and the sum
  $\sum_p$ runs over all $3!=6$ permutations of $\omega_a, \omega_b,
  \omega_c$.

  %-------------------------------------------------------
  %-------------------------------------------------------
  %-------------------------------------------------------
  \section{Expansion in Many-Electron Eigenstates}
  %-------------------------------------------------------
  %-------------------------------------------------------
  %-------------------------------------------------------

  Expressing the two-point correlation function
  Eq. (\ref{eq:sdefusingj}) using Eq. (\ref{eq:polarizationdef}) we
  get
  %-------------------------------------------------------
  \begin{equation}
    J(t) = \sum_n \langle B_n(t) B_n^{\dagger}(0) \rangle
  \end{equation}
  %-------------------------------------------------------- 
  Similarly, upon the substitution of Eq. (\ref{eq:polarizationdef})
  in Eq. (\ref{eq:fourpointCfunction}) we find three contributions $F
  = F_1 + F_2 + F_3$ where
  %--------------------------------------------------------
  \begin{eqnarray}\label{corrfuncdef}
    F_1(\tau_4, \tau_3, \tau_2, \tau_1) &=& \sum_{n,m}\langle B_m(\tau_4)
    B_m^{\dagger}(\tau_3) B_n(\tau_2) B_n^{\dagger}(\tau_1) \rangle, \nonumber \\
     F_2(\tau_4, \tau_3, \tau_2, \tau_1) &=& \sum_{n \neq m}\langle B_n(\tau_4)
     B_m(\tau_3) B_m^{\dagger}(\tau_2) B_n^{\dagger}(\tau_1) \rangle, \nonumber \\
     F_3(\tau_4, \tau_3, \tau_2, \tau_1) &=& \sum_{n \neq m}\langle B_m(\tau_4)
     B_n(\tau_3) B_m^{\dagger}(\tau_2) B_n^{\dagger}(\tau_1) \rangle.
  \end{eqnarray}
  %--------------------------------------------------------

  $F_{1}$ only contains transitions to and from the ground state $g
  \rightarrow n \rightarrow g \rightarrow m \rightarrow g$, and
  depends on either one ($n=m$) or two ($n\neq m$) core holes. Its
  sensitivity to correlations between core holes stems from an
  interference between two pathways that lead to the same valence
  electron-hole pair via the two possible intermediate channels (core
  hole on $n$ or $m$).  At no point along the path do we have a state
  with two core-holes existing simultaneously. $F_2$ and $F_3$, in
  contrast, are intrinsically cooperative since they also include
  transitions among the excited states $g \rightarrow n \rightarrow nm
  \rightarrow m \rightarrow g$ and depend on two-core-hole (two
  exciton) states.  This may best be seen using the Liouville space
  pathways (Eq. (\ref{eq:RandF})) displayed in Fig.
  \ref{fig:feynmandiag}.
  
  The evaluation of these matrix elements requires the many-electron
  wavefunctions $\vert \psi_{\nu}^N \rangle$ of the original molecule
  with N valence electrons, where $\nu = g$ is the ground state and
  $\nu = e, f \dots$ are valence excited states. In addition we need
  the valence $N+1$ electron wavefunctions calculated in the presence
  of the $n$th core hole, $\vert \psi_{n,\nu}^{N+1}\rangle$ and $N+2$
  electron wavefunctions calculated in the presence of two core holes
  at $n$ and $m$, $\vert \psi_{nm,\nu}^{N+2} \rangle$.  The
  corresponding energies will be denoted $E_{\nu}^{N}$,
  $E_{n,\nu}^{N+1} + \Omega_n$ and $E_{nm,\nu}^{N+2} + \Omega_n +
  \Omega_m$ respectively. Here $\Omega_n$ is the core-hole excitation
  energy whereas $E_{\nu}$ is the energy associated with valence
  electrons. Expanding in these states, the four point correlation
  functions (Eqn. (\ref{corrfuncdef})) assume the form
  \begin{eqnarray} \label{eq:sumoverstates}
    F_1(\tau_4, \tau_3, \tau_2, \tau_1) &=& \sum_{nm}\sum_{\nu_1\nu_2\nu_3} \langle \psi_{g}^{N} \vert
    B_m(\tau_4) \vert \psi_{m,\nu_3}^{N+1}
    \rangle
    \langle \psi_{m,\nu_3}^{N+1} \vert
    B_m^{\dagger}(\tau_3) \vert \psi_{\nu_2}^{N}
    \rangle \nonumber \\
    &\times& 
    \langle \psi_{\nu_2}^{N} \vert
    B_n(\tau_2) \vert \psi_{n,\nu_1}^{N+1}
    \rangle
    \langle \psi_{n,\nu_1}^{N+1} \vert
    B_n^{\dagger}(\tau_1) \vert \psi_{g}^{N}
    \rangle \nonumber\\
    F_2(\tau_4, \tau_3, \tau_2, \tau_1) &=& \sum_{n \neq m}\sum_{\nu_1\nu_2\nu_3} \langle \psi_{g}^{N} \vert
    B_n(\tau_4) \vert \psi_{n,\nu_3}^{N+1}
    \rangle
    \langle \psi_{n,\nu_3}^{N+1} \vert
    B_m(\tau_3) \vert \psi_{nm,\nu_2}^{N+2}
    \rangle \nonumber \\
    &\times&
    \langle \psi_{nm,\nu_2}^{N+2} \vert
    B_m^{\dagger}(\tau_2) \vert \psi_{n,\nu_1}^{N+1}
    \rangle
    \langle \psi_{n,\nu_1}^{N+1} \vert
    B_n^{\dagger}(\tau_1) \vert \psi_{g}^{N}
    \rangle \nonumber\\
    F_3(\tau_4, \tau_3, \tau_2, \tau_1) &=& \sum_{n \neq m}\sum_{\nu_1\nu_2\nu_3} \langle \psi_{g}^{N} \vert
    B_m(\tau_4) \vert \psi_{m,\nu_3}^{N+1}
    \rangle
    \langle \psi_{m,\nu_3}^{N+1} \vert
    B_n(\tau_3) \vert \psi_{nm,\nu_2}^{N+2}
    \rangle \nonumber \\
    &\times& 
    \langle \psi_{nm,\nu_2}^{N+2} \vert
    B_m^{\dagger}(\tau_2) \vert \psi_{n,\nu_1}^{N+1}
    \rangle
    \langle \psi_{n,\nu_1}^{N+1} \vert
    B_n^{\dagger}(\tau_1) \vert \psi_{g}^{N}
    \rangle \nonumber\\
  \end{eqnarray}

  These can be expressed as
  \begin{eqnarray}
    F_1(\tau_4, \tau_3, \tau_2, \tau_1) &=& \sum_{nm}\textrm{exp}\Big( -i
    \Omega_m \tau_{43} - i\Omega_n \tau_{21}\Big) \sum_{\nu_1 \nu_2
    \nu_3} \textrm{exp}\Big[ -i E_{m,\nu_3}^{N+1} \tau_{43} - i
    E_{\nu_2}^N \tau_{32} - i E_{n,\nu_1}^{N+1} \tau_{21}\Big]
    \nonumber\\
    &\times&
    \langle \psi_{g}^{N} \vert
    B_m \vert \psi_{m,\nu_3}^{N+1}
    \rangle
    \langle \psi_{m,\nu_3}^{N+1} \vert
    B_m^{\dagger} \vert \psi_{\nu_2}^{N}
    \rangle
    \langle \psi_{\nu_2}^{N} \vert
    B_n \vert \psi_{n,\nu_1}^{N+1}
    \rangle
    \langle \psi_{n,\nu_1}^{N+1} \vert
    B_n^{\dagger} \vert \psi_{g}^{N}
    \rangle \nonumber\\
    F_2(\tau_4, \tau_3, \tau_2, \tau_1) &=& \sum_{n \neq m}\textrm{exp}\Big( -i
    \Omega_m \tau_{32} - i\Omega_n \tau_{41}\Big) \sum_{\nu_1 \nu_2
    \nu_3} \textrm{exp}\Big[ -i E_{n,\nu_3}^{N+1} \tau_{43} - i
    E_{nm,\nu_2}^{N+2} \tau_{32} - i E_{n,\nu_1}^{N+1} \tau_{21}\Big]
    \nonumber\\
    &\times& \langle \psi_{g}^{N} \vert
    B_n \vert \psi_{n,\nu_3}^{N+1}
    \rangle
    \langle \psi_{n,\nu_3}^{N+1} \vert
    B_m \vert \psi_{nm,\nu_2}^{N+2} \rangle
    \langle \psi_{nm,\nu_2}^{N+2} \vert
    B_m^{\dagger} \vert \psi_{n,\nu_1}^{N+1}
    \rangle
    \langle \psi_{n,\nu_1}^{N+1} \vert
    B_n^{\dagger} \vert \psi_{g}^{N}
    \rangle \nonumber\\
    F_3(\tau_4, \tau_3, \tau_2, \tau_1) &=& \sum_{n \neq m}\textrm{exp}\Big( -i
    \Omega_m \tau_{42} - i\Omega_n \tau_{31}\Big)\sum_{\nu_1 \nu_2
    \nu_3} \textrm{exp}\Big[ -i E_{m,\nu_3}^{N+1} \tau_{43} - i
    E_{nm,\nu_2}^{N+2} \tau_{32} - i E_{n,\nu_1}^{N+1} \tau_{21}\Big]
    \nonumber\\
    &\times& \langle \psi_{g}^{N} \vert
    B_m \vert \psi_{m,\nu_3}^{N+1}
    \rangle
    \langle \psi_{m,\nu_3}^{N+1} \vert
    B_n \vert \psi_{nm,\nu_2}^{N+2} \rangle
    \langle \psi_{nm,\nu_2}^{N+2} \vert
    B_m^{\dagger} \vert \psi_{n,\nu_1}^{N+1}
    \rangle
    \langle \psi_{n,\nu_1}^{N+1} \vert
    B_n^{\dagger} \vert \psi_{g}^{N}
    \rangle \nonumber\\
  \end{eqnarray}
  where $\tau_{ij} \equiv \tau_{i}-\tau_{j}$ and $i, j =1,2,3,4$. For
  the linear response we have
  \begin{equation}
    J(\tau) = \sum_{n,\nu} \vert \langle \psi_{n,\nu}^{N+1} \vert
    B_n^{\dagger} \vert \psi_g^N \rangle \vert^2 \textrm{exp}(-i
    \Omega_n \tau - i E_{n,\nu}^{N+1} \tau )
  \end{equation}
  A finite core-hole lifetime can be added by setting $\Omega_m
  \rightarrow \Omega_m - \frac{i}{2}\gamma_m$, $\Omega_n \rightarrow
  \Omega_n - \frac{i}{2}\gamma_n$. $\gamma$ provides a time window for
  the experiment.  Typically it is $\sim 0.375$eV which corresponds to
  $\sim 10$~fsec window.  Only higher frequencies and faster processes
  than this window can be probed by resonant x-ray
  techniques. Information about multiple core-hole dynamics can be
  also extracted from frequency domain x-ray four wave
  mixing.\cite{mukpaper433}

  %---------------------------------------------------------
  %---------------------------------------------------------
  %---------------------------------------------------------
  \section{Coherent Multidimensional Signals}
  %---------------------------------------------------------
  %---------------------------------------------------------
  %---------------------------------------------------------
  
  We consider a sequence of x-ray pulses
  (Fig. \ref{fig:pulsesequence}), whose electric field is given by:
  %---------------------------------------------------------
  \begin{eqnarray}\label{eq:optlaukas}
    E(\mathbf{r},t) =\sum_{j=1}^{4}
    E_{j}(t)\exp \left( i\mathbf{k}_j
      \mathbf{r}-i\omega_j \tau \right)  + \textrm{c.c.}
  \end{eqnarray}
  %---------------------------------------------------------
  Here $E_{j} (t)$ is the slowly--varying complex envelope
  function of pulse $j$ with carrier frequency $\omega_j$ and
  wavevector $\mathbf{k}_j$. $c. c.$ denotes the complex
  conjugate. Most generally, a third-order process requires four
  external fields: three ($j$=1, 2, 3) interact with the system and
  the fourth, heterodyne, field ($j$=4) is used for the detection.

  To calculate the signals we expand the nonlinear polarization in
  $\mathbf{k}$ space:
  %----------------------------------------------------------------------
  \begin{equation}
    P_{}^{(3)}(\mathbf{r},t) = \sum_s P_{s}^{(3)}(t)
    \exp(i\mathbf{k}_s \mathbf{r}),
  \end{equation}
  %-------------------------------------------------------------
  where the possible wavevectors are $\mathbf{k}_s= \pm \mathbf{k}_1
  \pm \mathbf{k}_2 \pm \mathbf{k}_3$.  

  We shall consider well-separated pulses where pulse 1 comes first,
  followed by 2 and finally 3.\ref{fig:pulsesequence} Three signals
  are possible for our model, $\mathbf{k}_I = -\mathbf{k}_1
  +\mathbf{k}_2 + \mathbf{k}_3$, $\mathbf{k}_{II} = \mathbf{k}_1 -
  \mathbf{k}_2 + \mathbf{k}_3$, and $\mathbf{k}_{III} = \mathbf{k}_1 +
  \mathbf{k}_2 - \mathbf{k}_3$.  The polarizations responsible for the
  these signals, obtained by invoking the RWA (i.e. neglecting highly
  oscillatory off-resonant terms), are given by

  \begin{eqnarray}\label{eq:polarizations}
    P_I^{(3)}(t) &=& \int_{-\infty}^t d\tau_3 \int_{-\infty}^{\tau_3} d\tau_2 \int_{-\infty}^{\tau_2} d\tau_1 \nonumber \\
    && \Big[
    F_1(\tau_1,\tau_2,\tau_4,\tau_3) +
    F_1(\tau_1,\tau_3,\tau_4,\tau_2) -
    F_2(\tau_1,\tau_4,\tau_3,\tau_2) -
    F_3(\tau_1,\tau_4,\tau_3,\tau_2) \Big] \nonumber \\
    &\times& E_1^*(\tau_1) E_2(\tau_2)
    E_3(\tau_3) \nonumber\\
    P_{II}^{(3)}(t) &=& \int_{-\infty}^t d\tau_3 \int_{-\infty}^{\tau_3} d\tau_2 \int_{-\infty}^{\tau_2} d\tau_1 \nonumber \\
    && \Big[
    F_1(\tau_2,\tau_3,\tau_4,\tau_1) +
    F_1(\tau_4,\tau_3,\tau_2,\tau_1) -
    F_2(\tau_2,\tau_4,\tau_3,\tau_1) -
    F_3(\tau_2,\tau_4,\tau_3,\tau_1) \Big] \nonumber \\
    &\times& E_1(\tau_1) E_2^*(\tau_2)
    E_3(\tau_3) \nonumber\\
    P_{III}^{(3)}(t) &=& \int_{-\infty}^t d\tau_3 \int_{-\infty}^{\tau_3} d\tau_2 \int_{-\infty}^{\tau_2} d\tau_1 \nonumber \\
    && \Big[
    F_2(\tau_4,\tau_3,\tau_2,\tau_1) +
    F_3(\tau_4,\tau_3,\tau_2,\tau_1) -
    F_3(\tau_3,\tau_4,\tau_2,\tau_1) -
    F_2(\tau_3,\tau_4,\tau_2,\tau_1) \Big] \nonumber \\
    &\times& E_1(\tau_1) E_2(\tau_2)
    E_3^*(\tau_3) \nonumber\\
  \end{eqnarray}
  
  For very short (impulsive) pulses, we can eliminate the time
  integrations and simply set $\tau_1 = t-t_3-t_2-t_1$, $\tau_2 =
  t-t_3-t_2$, $\tau_3 = t- t_3$, and $\tau_4 = t$. $P_j^{(3)}$ will
  then depend parametrically on the three time delays $t_1$, $t_2$,
  and $t_3$ (Fig. \ref{fig:pulsesequence}). 

  The physical processes underlying each of these signals can be
  understood by using the Feynman diagrams shown in
  Fig. (\ref{fig:feynmandiag}) \cite{mukbook} which depict the
  evolution of the valence electronic density matrix in the course of
  the nonlinear process.  The two vertical lines in the diagram
  represent the ket and the bra (time goes from the bottom to the
  top), while arrows represent interactions with the laser pulses.  A
  coherence($\vert g \rangle \langle m \vert$ or $\vert m \rangle
  \langle g \vert$) is created by the first pulse. The second pulse
  takes the system either to a single exciton population ($\vert n
  \rangle \langle n \vert$), coherence ($\vert n \rangle \langle m
  \vert$), or to a two exciton coherence ($\vert n m \rangle \langle g
  \vert$). The population and the coherence evolutions can then be
  probed by holding the second delay time, $t_2$ fixed.  The third
  pulse creates coherences either between the ground and one-exciton
  states or between one- and two-exciton states.  The four diagrams
  contributing to $\mathbf{k}_I$ are shown in the left column. In all
  diagrams the density matrix represents a single-quantum coherence
  $\vert g \rangle \langle n \vert$ between the ground state and the
  singly excited state during $t_{1}$. During $t_{3}$ it is either in
  the conjugate coherence $\vert m \rangle \langle g \vert$ (a and b),
  or in a coherence between the one and two exciton manifolds $\vert
  mn \rangle \langle n \vert$ (c and d). $\mathbf{k}_{III}$ is
  similarly described by the four diagrams ((i),(j),(k) and (l)) and
  shows double-quantum coherences between ground state and the
  two--exciton band $\vert nm \rangle \langle n \vert$ during the
  $t_2$ interval.  During $t_{3}$ it has a single quantum coherence
  $\vert n\rangle \langle g \vert$ (i and j) and $\vert nm \rangle
  \langle n \vert$ (k and l). $\mathbf{k}_{I}$, known as the photon
  echo technique, can improve the resolution by eliminating certain
  types of inhomogeneous broadening.  $\mathbf{k}_{III}$ carries
  direct information regarding the coherence between the two exciton states
  and the ground state (double quantum coherence) so that its spectral
  bandwidth is doubled.\cite{mukbook, weipaper}
  
  Note that the absolute time arguments of $F$ in
  Eq. (\ref{eq:polarizations}) are not time ordered.  However they are
  ordered on the Keldysh loop shown in Fig. (\ref{fig:keldyshloop}) in
  the following sense: for each diagram we can start in the bottom
  right, move up on the bra line and then down on the ket line.  This
  will give the order of the $\tau$ arguments for each of the 12 terms
  in Eq. (\ref{eq:polarizations}).

  Within the slowly varying amplitude approximation the signal field
  is proportional to the polarization, $E_s(t) \propto
  iP_s^{(n)}(t)$\cite{mukbook}. The simplest detection
  measures the time integrated signal field intensity, and the third
  order signal in the $\mathbf{k}_s$ direction is given by:
  %--------------------------------------------------------------
  \begin{equation}
    I_{hom}(t_{1},t_{2}) =\int_{-\infty}^{+\infty} |P^{(3)}_{s}(t)|^2
    dt.
  \end{equation}
  %--------------------------------------------------------------
  This is known as the homodyne detection mode. Additional time
  resolution may be achieved by time gating which yields the absolute
  value of polarization itself $I_{hom}(t_{1},t_{2},t_{3}) =
  |P^{(3)}_{s}(t)|^2$.

  In heterodyne detection the generated field $E_s(t)$ is mixed with a
  fourth field, ${E}_4(t)$ which has the same wavevector and the
  heretodyne signal is given by:
  %--------------------------------------------------------------
  \begin{equation}\label{eq:heterodynesignal}
    I_{het} (t_{1},t_{2},t_{3})= Im \int_{-\infty}^{+\infty}
    {E}_4^{\ast}(t) P^{(3)}_s(t) dt .
  \end{equation}
  %--------------------------------------------------------------
  The time resolution is now determined by the heterodyne field, and
  the signal depends linearly rather than quadratically on
  $P^{(3)}_s(t)$. By choosing different phases of the heterodyne field
  it is possible to measure separately the real and the imaginary
  parts of the polarization.

  A mixed time/frequency representation of the signal may be useful to
  reveal correlations in the system. For example, one can display two
  dimensional $\omega_1 / \omega_3$ correlation plots for a fixed
  $t_2$.
  %--------------------------------------------------------------
  \begin{equation}\label{eq:hetero-mixed-signal}
    I_{het}(\omega_1,t_{2},\omega_3) = \int_{0}^{+\infty} \!\!\!
    \int_{0}^{+\infty}
    d t_1 d t_3
    I_{het}(t_1,t_{2},t_3)
    \exp(i\omega_1 t_1+i\omega_3 t_3).
  \end{equation}
  %--------------------------------------------------------------

  Time and frequency resolved signals and fields may be displayed
  using the Wigner spectrogram~\cite{khidekel,ciurdariu,schleich}:
  %--------------------------------------------------------------
  \begin{equation}\label{eq.wigner}
    W_s(t,\omega)=\int_{-\infty}^{+\infty}E_s^{\ast}(t-\tau/2)E_s(t+\tau/2)\exp(i\omega\tau)
    d\tau.
  \end{equation}
  %-------------------------------------------------------------- 
  The spectrogram directly shows what fraction of the field energy is
  contained in a given time and frequency window. Integrating over the
  frequencies gives the instantaneous field energy
  %--------------------------------------------------------------
  \begin{equation}\label{3.10a}
    \int_{-\infty}^{+\infty} W_s(t,\omega) d\omega=2\pi|E_s(t)|^2
  \end{equation}
  %--------------------------------------------------------------
  while integrating over the time gives the energy density spectrum
  %--------------------------------------------------------------
  \begin{equation}\label{3.11a}
    \int_{-\infty}^{+\infty} W_s(t,\omega)
    dt=|\mathcal{E}_s(\omega)|^2.
  \end{equation}
  %--------------------------------------------------------------
  The one dimensional projections of the spectrogram
  (Eqs. (\ref{3.10a}) and (\ref{3.11a})) are known as
  marginals.

  To express the heterodyne signal, Eqn. (\ref{eq:heterodynesignal}),
  in the Wigner representation we assume that the heterodyne field is
  a replica of one of the incoming fields in a nonlinear experiment,
  and expand the polarization to first order in this field:
  %--------------------------------------------------------------
  \begin{equation}
    P_s(t)=\int_{-\infty}^{+\infty} d\tau
    \tilde{S}^{(1)}(t,\tau)E_4(\tau).
  \end{equation}
  %--------------------------------------------------------------
  $\tilde{S}$ is a non-equilibrium correlation function of the system
  driven by all other fields. Defining the mixed time--frequency
  response function
  %--------------------------------------------------------------
  \begin{equation}
    \tilde{S}^{(1)}(t,\omega)=\int_{-\infty}^{+\infty}
    \tilde{S}^{(1)}(t+\tau/2,t-\tau/2) \exp(i\omega\tau) d\tau,
  \end{equation}
  %--------------------------------------------------------------
  the heterodyne signal assumes the form~\cite{khidekel,ciurdariu}:
  %--------------------------------------------------------------
  \begin{equation}\label{3.14a}
    I_{het}(t_{1},t_{2},t_{3})
    %(    \bar{\omega}_1,
    %       \bar{\tau}_1,
    %       \bar{\omega}_2,
    %       \bar{\tau}_{2},
    %       \bar{\omega}_3,
    %       \bar{\tau}_3,
    %       \bar{\omega}_4,
    %       \bar{\tau}_4
    % )
    =\int_{-\infty}^{+\infty}dt
    \int_{-\infty}^{+\infty}\frac{d\omega}{2\pi}W_4(t,\omega)
    \tilde{S}^{(1)}(t,\omega),
  \end{equation}
  %--------------------------------------------------------------
  where $W_4(t,\omega)$ is the heterodyne spectrogram, (Eq.~(\ref{eq.wigner})).

  Eq.~(\ref{3.14a}) is exact and holds for arbitrary field
  envelopes. For impulsive (very short) pulses the Wigner distribution is
  narrowly peaked at the time of heterodyne field $\bar{\tau}_4$ and
  Eq.~(\ref{3.14a}) reduces to
  %--------------------------------------------------------------
  \begin{equation}
    I_{het}(t_{1},t_{2},t_{3})
    %(    \bar{\omega}_1,
    %       \bar{\tau}_1,
    %       \bar{\omega}_2,
    %       \bar{\tau}_{2},
    %       \bar{\omega}_3,
    %       \bar{\tau}_3,
    %       \bar{\omega}_4,
    %       \bar{\tau}_4
    % )
    =\tilde{S}^{(1)}(\bar{\tau}_4,\bar{\tau}_4)\propto
    Im\{E_h^{\ast}(\bar{\tau}_4)P_s(\bar{\tau}_4)\}.
  \end{equation}
  %--------------------------------------------------------------
  In the other extreme of ideal frequency domain experiments the
  spectrogram is narrowly peaked around
  its carrier frequency $\bar{\omega}_4$ and
  %--------------------------------------------------------------
  \begin{equation}
    I_{het}(t_{1},t_{2},t_{3})
    %(    \bar{\omega}_1,
    %       \bar{\tau}_1,
    %       \bar{\omega}_2,
    %       \bar{\tau}_{2},
    %       \bar{\omega}_3,
    %       \bar{\tau}_3,
    %       \bar{\omega}_4,
    %       \bar{\tau}_4
    % )
    =\tilde{S}^{(1)}(\bar{\omega}_4,\bar{\omega}_4)\propto
    Im\big[\mathcal{E}_h^{\ast}(\bar{\omega}_4)P_s(\bar{\omega}_4)\big],
  \end{equation}
  %--------------------------------------------------------------
  where
  %--------------------------------------------------------------
  \begin{equation}
    \tilde{S}^{(1)}(\omega_1,\omega_2)=\int_{-\infty}^{+\infty}
    d\tau_1 d\tau_2 \tilde{S}^{(1)}(\tau_1,\tau_2)
    \exp(i\omega_1\tau_1+i\omega_2\tau_2) .
  \end{equation}
  %--------------------------------------------------------------

%----------------------------------------------------------------------
%----------------------------------------------------------------------
%----------------------------------------------------------------------
\section{Pump-Probe Spectroscopy}
%----------------------------------------------------------------------
%----------------------------------------------------------------------
%----------------------------------------------------------------------

  Pump-probe is the simplest third order technique and only requires
  two pulses. The signal defined as the difference absorption of the
  probe with and without the pump is related to the polarization at
  $\mathbf{k}_s = \mathbf{k}_1 + \mathbf{k}_2 - \mathbf{k}_1$
  originating from two interactions with the pump $(\omega_1,
  \mathbf{k}_1)$ and one with the probe $(\omega_2,
  \mathbf{k}_2)$. The probe serves as the heterodyne field since the
  signal is measured in the probe direction.  This technique may thus
  be viewed as self-heterodyne detection. This is an incoherent
  technique whereby the contributions of different molecules to the
  signal itself (rather than to its amplitude) are additive.

  We consider a \emph{sequential} pump probe signal induced by short,
  well-separated pulses where the pump comes first, followed by the
  probe with a delay time of $\tau$. The signal is obtained from
  Eqs. (\ref{eq:polarizations}) and (\ref{eq:heterodynesignal}) by
  combining the $\mathbf{k}_{I}$ and $\mathbf{k}_{II}$
  polarizations. We get $S_{pp} = S^A_{pp} + S^B_{pp}$, where

  \begin{eqnarray} \label{eq:sadef}
    S^{A}_{pp}(\omega_1, \omega_2; \tau) &=& \int_{-\infty}^{\infty} dt
    \int_{0}^{\infty} dt_3
    \int_{0}^{\infty} dt_2
    \int_{0}^{\infty} dt_1
    E_2^*(t-\tau + t_3) E_2(t - \tau) E_1^*(t - t_2) E_1 (t - t_2 - t_1) \nonumber \\
    &\times& \textrm{exp}[i \omega_2 t_3 + i \omega_1 t_1] \nonumber \\
    &\times& \Big[ F_1 (t - t_2 - t_1,  t - \tau + t_3, t - \tau,t - t_2) \nonumber \\
    &+& F_1(t - t_2 - t_1, t - t_2, t - \tau, t - \tau + t_3) \Big] \nonumber \\
    &+& \int_{-\infty}^{\infty} dt
    \int_{0}^{\infty} dt_3
    \int_{0}^{\infty} dt_2
    \int_{0}^{\infty} dt_1
    E_2^*(t-\tau + t_3) E_2(t - \tau) E_1(t - t_2) E_1^* (t - t_2 - t_1) \nonumber \\
    &\times& \textrm{exp}[i \omega_2 t_3 - i \omega_1 t_1] \nonumber \\
    &\times& \Big[ F_1 (t - t_2,  t - \tau + t_3, t - \tau,t - t_2-t_1) \nonumber \\
    &+& F_1(t - \tau,  t - \tau + t_3, t - t_2,t - t_2 - t_1)\Big] 
  \end{eqnarray}

  \begin{eqnarray} \label{eq:sbdef}
    S^{B}_{pp}(\omega_1, \omega_2; \tau) &=& \int_{-\infty}^{\infty} dt
    \int_{0}^{\infty} dt_3
    \int_{0}^{\infty} dt_2
    \int_{0}^{\infty} dt_1
    \textrm{exp}[-i \omega_2 t_3 + i \omega_1 t_1] \nonumber \\
    &\times& E_2(t+t_3-\tau) E_2^*(t-\tau) E_1^*(t - t_2) E_1 (t - t_2 - t_1) \nonumber \\
    &\times& \Big[ F_2 (t - t_2 - t_1,  t - \tau + t_3, t - \tau, t - t_2) \nonumber \\
    &+& F_3(t - t_2 - t_1,  t - \tau + t_3, t - \tau, t - t_2) \Big] \nonumber \\
    &+& \int_{-\infty}^{\infty} dt
    \int_{0}^{\infty} dt_3
    \int_{0}^{\infty} dt_2
    \int_{0}^{\infty} dt_1
    E_2(t+ t_3-\tau) E_2^*(t - \tau) E_1(t - t_2) E_1^* (t - t_2 - t_1) \nonumber \\
    &\times& \textrm{exp}[-i \omega_2 t_3 - i \omega_1 t_1] \nonumber \\
    &\times& \Big[ F_2 (t - t_2, t - \tau + t_3,  t - \tau, t - t_2-t_1) \nonumber \\
    &+& F_3(t - t_2,  t - \tau + t_3, t - \tau, t - t_2 - t_1) \Big] 
  \end{eqnarray}

  These terms can be separated into two negative contributions, ground
  state bleaching (GB) and stimulated emission (SE) and a positive
  path, excited state absorption (ESA). The corresponding Feynman
  diagrams are shown in Fig. \ref{fig:feynmandiagsecond}.

  Using the Wigner representation (Eq. (\ref{3.14a})), the pump probe
  signal can be expressed as an overlap integral of three functions:
  the pump spectrogram $W_1(t',\omega')$, the third order response
  function $S^{(3)}(t'',\omega'',t',\omega')$ and the probe
  spectrogram $W_2(t'',w'')$ \cite{khidekel,ciurdariu}:
  %-------------------------------------------------------------
  \begin{equation}
    I_{PP}(    \bar{\omega}_1,
    \bar{\tau}_1,
    \bar{\omega}_2,
    \bar{\tau}_{2},
    )=\int \!\!\!\int \!\!\!\int \!\!\!\int dt'dt''d\omega'd\omega''
    W_2(t'',w'') S^{(3)}(t'',\omega'',t',\omega')W_1(t',\omega').
  \end{equation}
  %----------------------------------------------------------------
  
  %---------------------------------------------------------
  %---------------------------------------------------------
  %---------------------------------------------------------
  \section{Fluorescence and Raman Spectroscopy}
  %---------------------------------------------------------
  %---------------------------------------------------------
  %---------------------------------------------------------
  Resonant x-ray emission is widely used in the study of core-hole
  transitions \cite{nordgren97, beyondoneelectronmodel, agrenreview99,
  nozieresandabrams, agren01}.  We consider a molecule driven by an
  x-ray field with a complex envelop $E(t)$ and carrier frequency
  $\omega_L$. We shall write the molecule-field interaction within
  the RWA in the form
  \begin{eqnarray}
    \label{interaction} H_{int} = E(t)
    B^{\dagger}\mbox{exp}\left(-i\omega_L t\right) + E^*(t) B
    \mbox{exp}\left(i\omega_L t\right).
  \end{eqnarray}
  $B^\dagger(B)$ are the exciton creation (annihilation) operators and
  the dipole operator is given by $\mu=$$B+B^\dagger$.
  The time and frequency resolved flourescence spectrum is given
  by \cite{cohentannoudjibook}
  \begin{eqnarray}
    S_{F} (\omega_L, \omega_S, t) = \int_{-\infty}^{\infty} d\tau \int_{-\infty}^{t}
    d\tau_1 \int_{-\infty}^{t + \tau} d\tau_2 \langle B(\tau_2)
    B^{\dagger} (t+\tau) B(t) B^{\dagger}(\tau_1) \rangle \nonumber \\
    E (\tau_2) E^*(\tau_1) \textrm{exp}\left[ i \omega_L(\tau_1 -
      \tau_2) - i \omega_S \tau \right]
  \end{eqnarray}

  We can divide the response into time-ordered contributions which
  separate the Liouville space pathways into Raman and fluorescence
  types\cite{mukbook}. The three pathways with $\tau>0$ and $\tau<0$
  are complex conjugates. This gives
  \begin{eqnarray}\label{sledef}
    S_{F} (\omega_L, \omega_S, t) &=& 2 \textrm{Re} \int_0^{\infty}
    dt_3 \int_0^{\infty} dt_2 \int_0^{\infty} dt_1 \nonumber \\
    && \Big[ E (t - t_1 - t_2
      - t_3) E^*(t - t_2 - t_3) \textrm{exp}(i \omega_L t_1 + i
      \omega_S t_3) F_1 (t_1, t_1 + t_2, t_1 + t_2 + t_3, 0)\nonumber \\
      &+& E^*(t - t_1 - t_2 - t_3) E (t - t_2 - t_3)
      \textrm{exp}(-i \omega_L t_1 + i \omega_S t_3) F_1 (0, t_1 +
      t_2, t_1 + t_2 + t_3, t_1) \nonumber\\
      &+& E^*(t - t_1 - t_2 - t_3) E (t - t_3) \textrm{exp}(-i
      \omega_L t_1 - i (\omega_L - \omega_S) t_2 + i\omega_S t_3)
      \nonumber\\
      &\times& F_1 (0, t_1, t_1 + t_2 + t_3, t_1 + t_2) \Big]
  \end{eqnarray}
  The three terms in Eqn. (\ref{sledef}) come from $R_1$, $R_2$ and
  $R_3$ respectively of Eqn. (\ref{eq:RandF}).  It is interesting to
  note that $S_{F}$ only depends on $F_1$;  four wave mixing signals
  also depend on $F_2$ and $F_3$, and therefore explore new regimes
  of Fock space not accessible by fluorescence.

  We now insert the complete basis of many-body states in
  Eq. (\ref{eq:sumoverstates}). We denote the transition frequency
  between states $| a \rangle$ and $| b\rangle$ as $\omega_{ab} \equiv
  \omega_a - \omega_b$ and $\Gamma_{ab} = \frac{1}{2}(\gamma_a +
  \gamma_b)+\hat{\Gamma}_{ab}$ is the corresponding dephasing
  rate. Here $\gamma_a$ is the inverse lifetime of state $\vert a
  \rangle$ and $\hat{\Gamma}_{ab}$ is the pure dephasing rate
  resulting from frequency fluctuations. Assuming a c.w. field
  ($E(t)=1$), we can carry out the time integrations and obtain
  \begin{eqnarray}
    \label{total-signal-new} S_F(\omega_S,\omega_L,t) =
    -\frac{2}{\hbar^4}\mbox{Re}\left(\frac{}{}S_{I}(\omega_S,\omega_L)+
    S_{II}(\omega_S,\omega_L)+
    S_{III}(\omega_S,\omega_L)\frac{}{}\right),
  \end{eqnarray}
  where
  \begin{eqnarray}
    \label{s11-2} S_{I}(\omega_S,\omega_L) = -i
	\sum_{mn}\!\sum_{\nu_1,\nu_2,\nu_3}\! \frac{ \left\langle
      \psi^N_g|B_m|\psi^{N+1}_{m,\nu_1}\right\rangle 
      \left\langle \psi^{N+1}_{m,\nu_1}|B^{\dag}_m|
      \psi^N_{\nu_2}\right\rangle
      \left\langle \psi^N_{\nu_2}|B_n|\psi^{N+1}_{n,\nu_3}\right\rangle
     \left\langle \psi^{N+1}_{n,\nu_3}|B^{\dag}_n|\psi^N_g\right\rangle}
    {(\omega_{\nu_1\nu_2}-\omega_S+i\Gamma_{\nu_1\nu_2})(\omega_{\nu_1\nu_3}+i\Gamma_{\nu_1\nu_3})
	(\omega_{\nu_1g}-\omega_L+i\Gamma_{\nu_1g})} \nonumber \\
  \end{eqnarray}
  \begin{eqnarray}
    \label{s12-new} S_{II}(\omega_S,\omega_L) = -i  
    \sum_{mn}\!\sum_{\nu_1,\nu_2,\nu_3}\! \frac{ \left\langle
      \psi^N_g|B_m|\psi^{N+1}_{m,\nu_1}\right\rangle
      \left\langle \psi^{N+1}_{m,\nu_1}|B^{\dag}_m|
      \psi^N_{\nu_2}\right\rangle
      \left\langle \psi^N_{\nu_2}|B_n|\psi^{N+1}_{n,\nu_3}\right\rangle
     \left\langle \psi^{N+1}_{n,\nu_3}|B^{\dag}_n|\psi^N_g\right\rangle}
{(\omega_{\nu_1\nu_2}-\omega_S+i\Gamma_{\nu_1\nu_2})(\omega_{\nu_1\nu_3}+i\Gamma_{\nu_1\nu_3})
	(\omega_L-\omega_{\nu_3g}+i\Gamma_{\nu_3g})} \nonumber \\
  \end{eqnarray}
  \begin{eqnarray}
    \label{s13-new} S_{III}(\omega_S,\omega_L) = -i \sum_{mn}\!\sum_{\nu_1,\nu_2,\nu_3}\! \frac{ \left\langle
      \psi^N_g|B_m|\psi^{N+1}_{m,\nu_1}\right\rangle
      \left\langle \psi^{N+1}_{m,\nu_1}|B^{\dag}_m|
      \psi^N_{\nu_2}\right\rangle
      \left\langle \psi^N_{\nu_2}|B_n|\psi^{N+1}_{n,\nu_3}\right\rangle
     \left\langle \psi^{N+1}_{n,\nu_3}|B^{\dag}_n|\psi^N_g\right\rangle}
{(\omega_L-\omega_S-\omega_{\nu_2g}+i\Gamma_{\nu_2g})
	(\omega_{\nu_1\nu_2}-\omega_S+i\Gamma_{\nu_1\nu_2})(\omega_L-\omega_{\nu_3g}+i\Gamma_{\nu_3g})}
      . \nonumber \\
  \end{eqnarray}
  In the absence of dephasing, $\Gamma_{g\nu}=0, (\nu =
  \nu_1,\nu_2,\nu_3)$, these terms can be combined to yield
  \begin{eqnarray}
    S(\omega_S,\omega_L) &=&
    \frac{2\pi}{\hbar^4}\sum_{\nu_2}\left|\sum_{n}\sum_{\nu_1}\frac{\left\langle
      \psi^N_{\nu_2}|B_n|\psi^{N+1}_{n,\nu_1}\right\rangle
      \left\langle \psi^{N+1}_{n,\nu_1}|B^{\dag}_n|
      \psi^N_{g}\right\rangle}
{\omega_L-\omega_{g \nu_1}+i\eta}\right|^2  
  \delta(\omega_L-\omega_S-\omega_{g \nu_2})
  \end{eqnarray}
  where $\eta$ is an infinitesimal positive number. This is the
  standard expression for spontaneous emission spectra\cite{mukbook}.

%----------------------------------------------------------------------
%----------------------------------------------------------------------
%----------------------------------------------------------------------
  \section{The Electron-Boson Model}
%----------------------------------------------------------------------
%----------------------------------------------------------------------
%----------------------------------------------------------------------

  So far we derived exact expressions for the multipoint correlation
  functions in terms of the many-electron states.  A much simpler
  description can be obtained by replacing the valence excitations by a
  boson bath described by the operators $a_s$, $a_s^{\dagger}$ and
  adding the electron-boson model Hamiltonian \cite{hedin99,
  langreth70, lundqvist67, overhauser71, hedin85, hedin98}
  \begin{equation}\label{eq:ebmham}
    H = \sum_k \epsilon_k c_k^{\dagger} c_k + \sum_s \omega_s
    a_s^{\dagger} a_s + \sum_{s k } V_{k s} (a_s + a_s^{\dagger})
    c_k^{\dagger} c_{k}.
  \end{equation}
  The first two terms represent the reference Hamiltonian for the
  noninteracting valence and core electrons. The last term is the
  potential induced by the $k$'th core hole which causes a linear
  displacement of the bath modes.
  \begin{equation}\label{eq:ebmcorepot}
    U_k = \sum_s V_{ks} (a_s + a_s^{\dagger}), \qquad k=n,m
  \end{equation}

  Each time we act with $B^{\dagger}$ we add a valence electron.  If
  the valence system is very large (e.g. the electron gas or a large
  metal nanoparticle) we can ignore the effect of the additional
  electron and only consider the added core hole potential. In this
  case we can set $B^{\dagger} = c^{\dagger}$, $B=c$ and the
  polarization operator becomes
  %----------------------------------------------------------
  \begin{equation}\label{eq:PolOpMolecular}
    \hat{\mathbf{P}} = \sum_k (c_k + c_k^{\dagger} ).
  \end{equation}
  %----------------------------------------------------------
  We then have
  \begin{eqnarray}
    B_n^{\dagger}(\tau) &=& \textrm{exp}(i H_n \tau) \textrm{exp}(-i H_o \tau) \nonumber\\
    &=& \textrm{exp}_-\left[i \int_0^{\tau}d\tau U_n(\tau) \right],
  \end{eqnarray}
  where $H_o$ consists of the first two terms in
  Eqn. (\ref{eq:ebmham}) so that $U_n \equiv H_n - H_o$ is the
  \emph{fluctuating potential} caused by the $n$th core hole.
  \begin{equation}
    U_n(\tau) = \textrm{exp}(i H_o \tau) U_n \textrm{exp}(-i H_o \tau)
  \end{equation}
  Similarly
  \begin{equation}
    B_n(\tau) = \textrm{exp}_{+} \left[ -i \int_0^{\tau} d\tau U_n(\tau) \right]
  \end{equation}
  $\textrm{exp}_{+(-)}$ refer to positive(negative) time-ordered
  exponentials.  The multipoint correlation functions may then be
  calculated using the second order cumulant
  expansion.\cite{dariusreview,mukpaper, almbladh04}

  This is formally analogous to a multilevel molecular aggregate
  Hamiltonian with diagonal energy fluctuations whose nonlinear
  response was calculated in \cite{dariusreview,mukpaper,mukbook}.
  Assuming that the transition dipole operator does not depend on the
  bath coordinates, we obtain
  %----------------------------------------------------------------
  \begin{equation}
    J(\tau) = \sum_n \Big \langle \textrm{exp}(-i \int_0^\tau d\tau_1
    U_n (\tau_1) \Big \rangle
  \end{equation}
  \begin{equation}
    J(\tau)= \sum_{n} |\mu_{gn}|^2
    \exp[ -i\Omega_{n} \tau - \frac{1}{2}g_{nn}(\tau)
    ],
  \end{equation}
  %----------------------------------------------------------------
  $\mathbf{\mu}_{ag}$ is the transition dipole moment between the
  ground state and state $a$ and $\mathbf{\mu}_{ab}$ is the transition
  dipole moment between excited states $a$ and $b$. The line
  broadening function $g_{nm}$ is associated with energy level
  fluctuations:
  \begin{eqnarray}\label{primeg}
    g_{ab}(\tau)\equiv \frac{1}{2} \int_{0}^{\tau} d\tau_{1}
    \int_{0}^{\tau_{1}} d\tau_{2}\,\,\left[\langle U_{a}(\tau_{1})
    U_{b}(\tau_{2})\rangle + \langle U_{b}(\tau_{1})
    U_{a}(\tau_{2})\rangle \right].
  \end{eqnarray}
  %--------------------------------------------------------------
  %---------------------------------------------------------

  The four-point correlation functions may be calculated by starting
  with~\cite{mukpaper},

  \begin{eqnarray}
    F_1(\tau_4, \tau_3, \tau_2, \tau_1) &=& \sum_{n,m}\Big\langle
    \textrm{exp}_+\left(-i \int_0^{\tau_4} U_m(\tau) d\tau \right)
    \textrm{exp}_-\left(i \int_0^{\tau_3} U_m(\tau) d\tau \right) \nonumber\\
    &\times& \textrm{exp}_+\left(-i \int_0^{\tau_2} U_n(\tau) d\tau \right)
    \textrm{exp}_-\left(i \int_0^{\tau_1} U_n(\tau) d\tau \right)
    \Big \rangle \nonumber \\
    F_2(\tau_4, \tau_3, \tau_2, \tau_1) &=& \sum_{n \neq m}\Big\langle
    \textrm{exp}_+\left(-i \int_0^{\tau_4} U_n(\tau) d\tau \right)
    \textrm{exp}_+\left(-i \int_0^{\tau_3} U_m(\tau) d\tau \right) \nonumber\\
    &\times& \textrm{exp}_-\left(i \int_0^{\tau_2} U_m(\tau) d\tau \right)
    \textrm{exp}_-\left(i \int_0^{\tau_1} U_n(\tau) d\tau \right)
    \Big \rangle \nonumber \\
    F_3(\tau_4, \tau_3, \tau_2, \tau_1) &=& \sum_{n \neq m}\Big\langle
    \textrm{exp}_+\left(-i \int_0^{\tau_4} U_m(\tau) d\tau \right)
    \textrm{exp}_+\left(-i \int_0^{\tau_3} U_n(\tau) d\tau \right) \nonumber\\
    &\times& \textrm{exp}_-\left(i \int_0^{\tau_2} U_m(\tau) d\tau \right)
    \textrm{exp}_-\left(i \int_0^{\tau_1} U_n(\tau) d\tau \right)
    \Big \rangle
  \end{eqnarray}

  Performing the cumulant expansion to second order,\cite{ravi,meiereee,dariusreview} we get
  
  %---------------------------------------------------------
  \begin{eqnarray}\label{eq:f_simplified}
    F_1(\tau_4,\tau_3,\tau_2,\tau_1)&= &\sum_{n,m}
    \mu_{gm}
      \mu_{mg}
      \mu_{gn}
      \mu_{ng}\times \nonumber\\
      && \exp [ -i\Omega_{m} \tau_{43} -i\Omega_{n} \tau_{21}
        -f_1(\tau_4,\tau_3,\tau_2,\tau_1) ] \nonumber\\
  \end{eqnarray}
  
  %---------------------------------------------------------
  with
  %--------------------------------------------------------------
  \begin{eqnarray}\label{eq:f1def}
    f_{1}(\tau_{1},\tau_{2},\tau_{3},\tau_{4})= & & g_{nn}(\tau_{21})+
    g_{mm}(\tau_{43})+ g_{nm}(\tau_{32})+ g_{nm}(\tau_{41})-
    \nonumber \\
    && -g_{nm}(\tau_{31})-g_{nm}(\tau_{42}).
  \end{eqnarray}
  
  \begin{eqnarray}
    F_2(\tau_4,\tau_3,\tau_2,\tau_1) &=& \sum_{n \neq m}
    \mu_{gn}
    \mu_{gm}
    \mu_{mg}
    \mu_{ng}\times \nonumber\\
    && \exp [ -i\Omega_{m}\tau_{32} - i\Omega_{n}\tau_{41}
      - f_2(\tau_4,\tau_3,\tau_2,\tau_1) ],
  \end{eqnarray}
  with
  \begin{eqnarray}\label{eq:f2def}
    f_{2}(\tau_{1},\tau_{2},\tau_{3},\tau_{4})= & & g_{nn}(\tau_{41})+
    g_{mm}(\tau_{32}) - g_{nm}(\tau_{21})
    \nonumber \\
    && + g_{nm}(\tau_{31}) - g_{mn}(\tau_{43}) +g_{mn}(\tau_{42}).
  \end{eqnarray}
  and  
  \begin{eqnarray}
    F_3(\tau_4,\tau_3,\tau_2,\tau_1) &=& \sum_{n \neq m}
    \mu_{gm}
    \mu_{gn}
    \mu_{mg}
    \mu_{ng}\times \nonumber\\
    && \exp [ -i\Omega_{m}\tau_{42} -i \Omega_{n}\tau_{31}
      -f_3(\tau_4,\tau_3,\tau_2,\tau_1) ],
  \end{eqnarray}
  with  
  \begin{equation}
    f_3(\tau_4,\tau_3,\tau_2,\tau_1) = g_{nn}(\tau_{31}) + g_{mm}(\tau_{42}) +
    g_{nm}(\tau_{32}) - g_{nm}(\tau_{21}) + g_{nm}(\tau_{41}) - g_{nm}(\tau_{43}).
  \end{equation}
  %--------------------------------------------------------------
   
  The fluctuating potentials enter the response through the spectral
  densities:
  %---------------------------------------------------------
  \begin{eqnarray}\label{totalspectraldensitydiag}
    {C''}_{ab} (\omega) \equiv -\frac{1}{2}
    \int_{-\infty}^{+\infty} dt \exp(i \omega t) \langle [
      U_{a}(t) , U_{b}(0)] \rangle,
  \end{eqnarray}
  %--------------------------------------------------------- 
  where the expectation value and the time evolution are taken with
  respect to the bath Hamiltonian. These contain all relevant
  information about the fluctuations necessary for computing the
  nonlinear response of the system. $g_{ab} (t)$ are directly related
  to the diagonal spectral densities of the bath
  ${C''}_{ab}(\omega)$~\cite{mukbook}:
  %---------------------------------------------------------
  \begin{eqnarray}
    g_{ab}(t) &=& \int_{-\infty}^{\infty} \frac{d \omega}{2 \pi}
    \frac{1-\cos(\omega t)}{\omega^2} \coth\left(\frac{\hbar
      \omega}{2k_B T}\right) \left[C^{''}_{ab}(\omega) +
      C^{''}_{ba}(\omega) \right]\nonumber \\
    &&+ i \int_{-\infty}^{\infty} \frac{d \omega}{2 \pi}
    \frac{\sin(\omega t) - \omega t}{\omega^2} \left[
    C^{''}_{ab}(\omega) + C^{''}_{ba}(\omega) \right]
  \end{eqnarray}
  %---------------------------------------------------------
  These spectral densities can be obtained from photooelectron
  spectroscopy or inelastic x-ray scattering.\cite{schulke} A simple model
  for the bath is given by the overdamped Brownian oscillator spectral
  density:\cite{mukbook}
  %------------------------------------------------------------------------
  \begin{equation}\label{eq:lorentzbathdiag}
    {C''}_{ab}(\omega)=2\lambda_{ab}\frac{\omega\Lambda_{ab}}{\omega^2+\Lambda_{ab}^2}.
  \end{equation}
  %------------------------------------------------------------------------

  $\lambda_{aa}$ represents the magnitude of fluctuations of the
  energy of state $a$, while $\lambda_{ab}$ represents the fluctuation
  of the coupling between states $a$ and $b$.  It can be observed in
  fluorescence as the time dependent Stokes shift. We further define
  the linewidth parameter $\Delta_{ab}^2 \equiv 2k_BT\lambda_{ab}$.
  Substituting the overdamped Brownian oscillator spectral density we
  get in the high temperature limit:
  %---------------------------------------------------------
  \begin{equation}
    g_{ab}(t)=2 \left( \frac{2T\lambda_{ab}} {\Lambda_{ab}^2}
    -i\frac{\lambda_{ab}} {\Lambda_{ab}} \right)
    (\exp(-\Lambda_{ab}|t|) +\Lambda_{ab} t-1).
  \end{equation}

  Two dimensionless parameters, $\eta$ and $\kappa$, can be used to
  characterize the model and classify different regimes of energy
  fluctuations. The first, $\eta$, defined by~\cite{ravi}
  $\Delta_{ab}^2 \equiv \eta_{ab}\Delta_{aa}\Delta_{bb}$, represents
  the correlation of fluctuation amplitudes. $-1\leq\eta_{ab}\leq
  1$. These may be anti-correlated ($\eta_{ab}=- 1$), uncorrelated
  ($\eta_{ab}=0$) and fully correlated ($\eta_{ab}= 1$). The second
  parameter, $\kappa_{ab} \equiv \Lambda_{ab}/\Delta_{ab}$, is the
  ratio of the inverse time--scale of the bath to the amplitude of the
  fluctuations. It controls the lineshape; in the slow bath limit
  ($\kappa_{ab}<1$) it has a Gaussian profile which gradually turns
  into a Lorentzian as $\kappa_{ab}$ is increased~\cite{mukbook,Kubo}.

  %--------------------------------------------------------------
  %--------------------------------------------------------------
  %--------------------------------------------------------------
  \section{Discussion}
  %--------------------------------------------------------------f
  %--------------------------------------------------------------
  %--------------------------------------------------------------

  Nonlinear core-hole spectroscopies could provide critical tests for
  the limitations of the electron-boson model. Even when the core
  holes are localized, the valence orbitals are usually delocalized
  and the third order response (Eqs. (\ref{eq:sumoverstates}))
  requires four sets of valence orbitals corresponding to $\vert
  \psi^{N} \rangle $, $\vert \psi_m^{N+1} \rangle$, $\vert
  \psi_n^{N+1} \rangle$ and $\vert \psi_{nm}^{N+2} \rangle$. New
  insights could be provided on electron dynamics in conjugated
  molecules,\cite{mukpaper433,agrenreview99} and mixed valence
  compounds.\cite{dexheimer00} In the EBM, the effect of both core
  holes on the boson system is additive, (Eqn. (\ref{eq:ebmcorepot})).
  This is why the response only depends on two collective coordinates
  (fluctuation potentials), $U_n$ and $U_m$.  It is possible to extend
  this model in various ways.  For example, if we add an extra
  coupling to Eqn. (\ref{eq:ebmham}) $H' = \sum_{nm} V_{nm,s} (a_s +
  a_s^{\dagger}) c_n^{\dagger} c_m^{\dagger} c_n c_m$, the dynamics
  will depend on a third collective coordinate, $U_{nm} \equiv
  \sum_{s} V_{nm,s} (a_s + a_s^{\dagger})$. The bath displacement when
  both $n$ and $m$ holes exist simultaneously will then be
  non-additive, $U_n + U_m + U_{nm}$. The nonlinear response for this
  model may be calculated using the general expressions for a
  multilevel system given in ref (\cite{dariusreview}).  Fluorescence
  will not depend on $U_{nm}$, since it does not affect $F_1$. The
  valence electrons generally behave as anharmonic oscillators and the
  Hamiltonian may contain higher order terms in $a_s$ and
  $a^{\dagger}_s$. Coherent nonlinear x-ray techniques may provide new
  information about these anharmonicities in the same way that
  infrared multidimensional techniques probe vibrational potential
  surfaces.\cite{smannreview} The signal generally depends on many
  pulse parameters and can be displayed by various types of
  multidimensional correlation plots. For example, displaying the
  signals as a function of the time delays $t_1$, $t_2$ and $t_3$
  provides a direct look at valence electron wavepackets. In the
  electron gas these probe charge density fluctuations. The third
  order techniques considered in this article offer numerous new
  probes for electron dynamics.  Complementary information is provided
  by second order techniques such as sum and difference frequency
  generation.\cite{mukpaper494} These can also be analyzed using the
  formalism presented in the article.\cite{mukpaper512}

  Nonlinear core hole spectroscopy provides new ways for probing the
  dynamics and response of electronic valence excitations by
  controlled attosecond switching of external potentials.  Femtosecond
  pulses introduced in the eighties allowed real time probing
  of nuclear motions. Attosecond x-ray pulses make it possible to
  watch electronic motions in real time. These techniques could help
  develop more realistic anharmonic Hamiltonians for the core-valence
  couplings, and test the validity of approximate model Hamiltonians.

  \begin{acknowledgments}
    The support of Chemical Sciences, Geosciences and Biosciences
    Division, Office of Basic Energy Sciences, Office of Science,
    U.S. Department of Energy is gratefully acknowledged. I wish to
    thank Daniel Healion and Rajan Pandey for their help in the
    preparation of the manuscript.
  \end{acknowledgments}

  \begin{figure}
    \begin{center} 
      \includegraphics[height=.75\textheight]{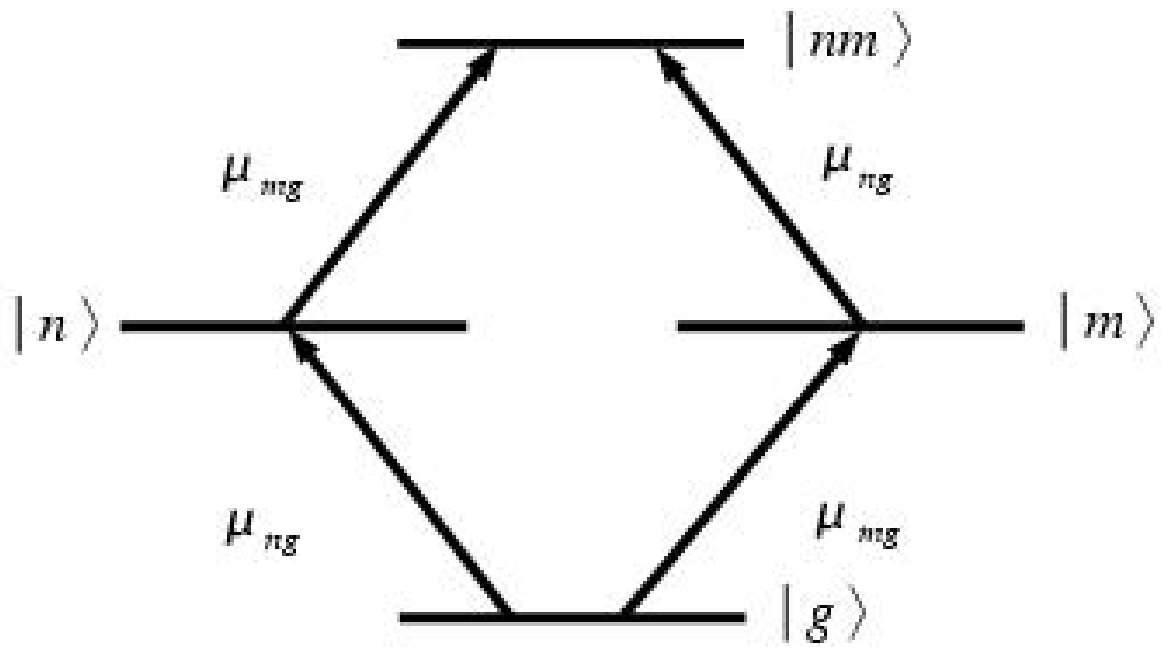}
      \caption{The core exciton states which enter the third order
	response.  $\vert g \rangle$ is the ground state, $\vert n
	\rangle$ and $\vert m \rangle$ are single core-hole states,
	and $\vert nm \rangle$ is a double core hole state.}
      \label{fig:levelscheme}
    \end{center}
  \end{figure}
  \begin{figure}  
    \includegraphics[height=.75\textheight]{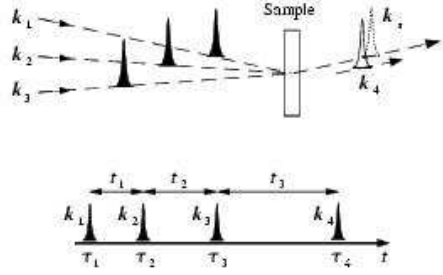}
    \caption{(Upper panel) Laser pulse sequence in a three-pulse,
      four-wave mixing experiment.  Three pulses $\mathbf{k}_1$,
      $\mathbf{k}_2$ and $\mathbf{k}_3$ create the nonlinear
      polarization in the sample, which generates the new x-ray field
      in the direction $\mathbf{k}_s = \pm \mathbf{k}_1 \pm \mathbf{k}_2
      \pm \mathbf{k}_3$. $\mathbf{k}_4 = \mathbf{k}_s$ is the heterodyne
      field. (Lower panel) The peak ordering and time intervals.}
    \label{fig:pulsesequence}
  \end{figure}    
  \begin{figure} 
    \includegraphics[height=.70 \textheight]{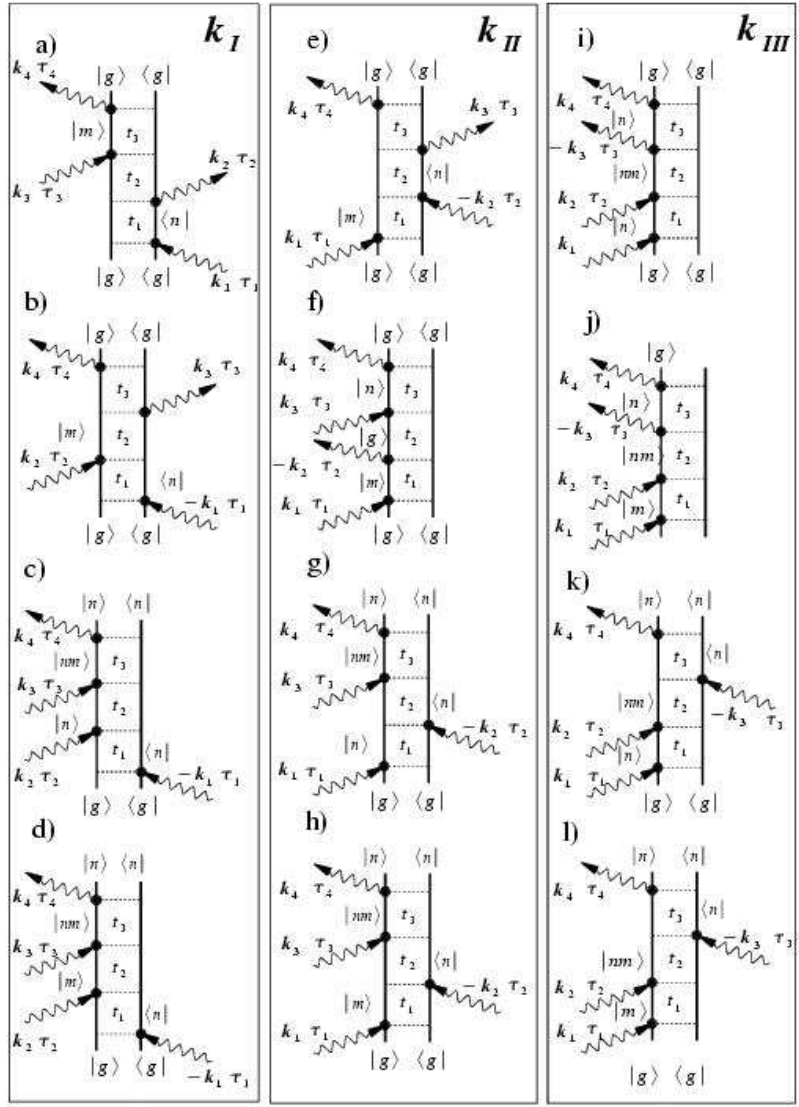}
    \caption{Double-sided Feynman diagams representing the
      Liouville-space pathways contributing to the third-order
      response in the rotating wave approximation. The level scheme is
      given in Fig. \ref{fig:levelscheme}.  Shown are the diagrams
      contributing to the four-wave mixing signal generated along the
      various possible directions: $\mathbf{k}_I = -\mathbf{k}_1 +
      \mathbf{k}_2 + \mathbf{k}_3$, $\mathbf{k}_{II} = \mathbf{k}_1 -
      \mathbf{k}_2 + \mathbf{k}_3$, and $\mathbf{k}_{III} =
      \mathbf{k}_1 + \mathbf{k}_2 - \mathbf{k}_3$. Diagrams (a), (b),
      (e), and (f) correspond to $F_1$ and only include one-exciton
      states. All other diagrams also involve two-exciton states. (i),
      (c), (g), and (k) correspond to $F_2$ and (j), (d), (h) and (l)
      represent $F_3$.}
    \label{fig:feynmandiag}
  \end{figure}
  \begin{figure}
    \includegraphics[height = 0.85 \textheight]{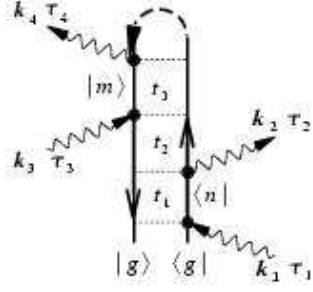}
    \caption{Each of the 12 non-time-ordered correlation function
    expressions for $F_j$ (Eq. (\ref{eq:polarizations})) may be
    obtained from the corresponding Feynman diagram
    (Fig. \ref{fig:feynmandiag}) using time ordering on a Keldysh
    loop. This is illustrated here for
    $\mathbf{k}_{I}$(a). ($F_1(\tau_1, \tau_2, \tau_4, \tau_3)$)}
    \label{fig:keldyshloop}
  \end{figure}
  \begin{figure} 
    \includegraphics[height=.75 \textheight]{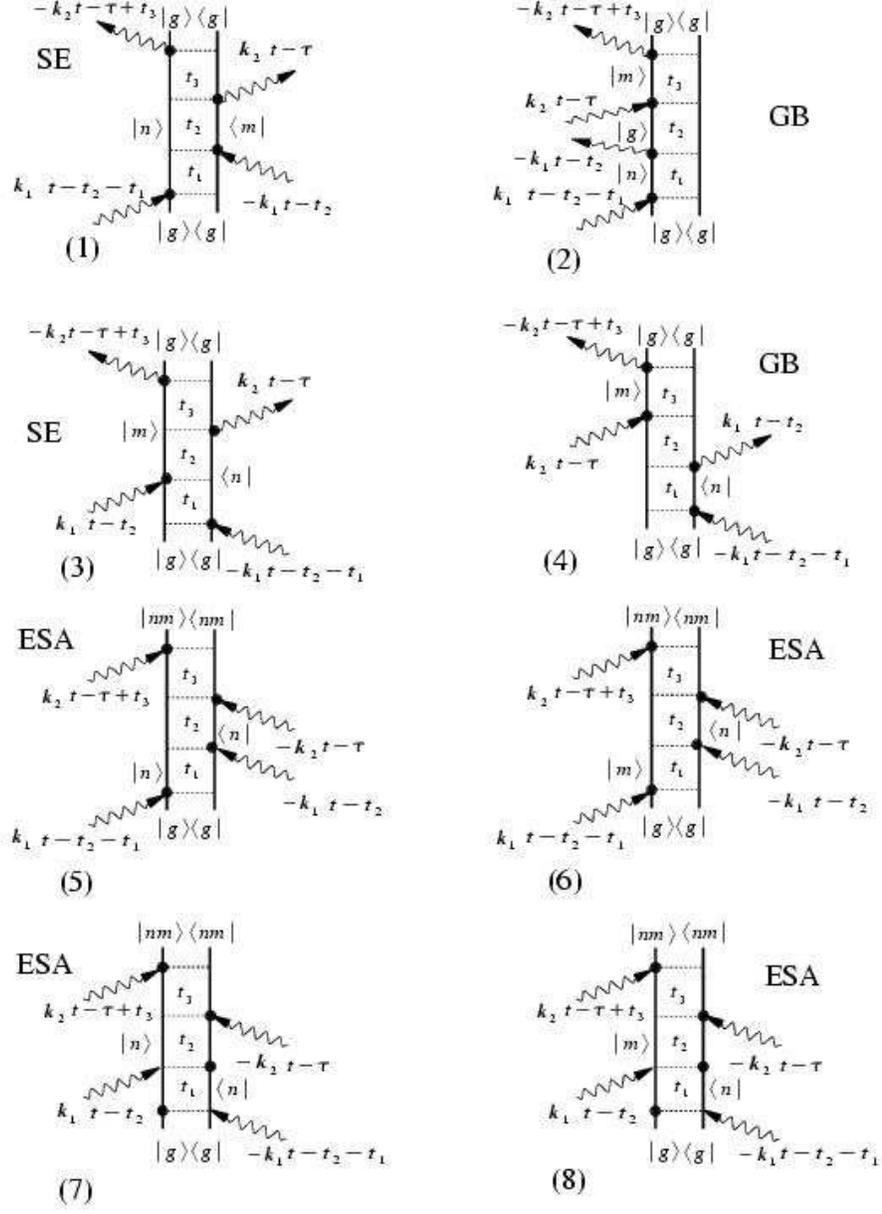}
    \caption{Feynman diagrams representing the eight contributions to
      the sequential pump-probe spectrum. Diagrams (1)-(8) correspond
      respectively to the eight terms in Eqs. (\ref{eq:sadef}) and
      (\ref{eq:sbdef}).  (1)-(4) represent the $F_1$ contributions
      (Eq. (\ref{eq:sadef})), which can be divided into ground state
      bleaching (GB) and stimulated emission (SE) type. (5)-(8) are
      the $F_2$ and $F_3$ contributions (Eq. (\ref{eq:sbdef})), which
      represent excited state absorption (ESA).}
    \label{fig:feynmandiagsecond}
  \end{figure}

\end{document}